# PDL as a Multi-Agent Strategy Logic

## Extensive Abstract[*]

Jan van Eijck
CWI and ILLC
Science Park 123
1098 XG Amsterdam, The Netherlands
jve@cwi.nl

## ABSTRACT

Propositional Dynamic Logic or PDL was invented as a logic for reasoning about regular programming constructs. We propose a new perspective on PDL as a multi-agent strategic logic (MASL). This logic for strategic reasoning has group strategies as first class citizens, and brings game logic closer to standard modal logic. We demonstrate that MASL can express key notions of game theory, social choice theory and voting theory in a natural way, we give a sound and complete proof system for MASL, and we show that MASL encodes coalition logic. Next, we extend the language to epistemic multi-agent strategic logic (EMASL), we give examples of what it can express, we propose to use it for posing new questions in epistemic social choice theory, and we give a calculus for reasoning about a natural class of epistemic game models. We end by listing avenues for future research and by tracing connections to a number of other logics for reasoning about strategies.

## Categories and Subject Descriptors

F.4.1 [**Mathematical Logic**]: Modal Logic; F.4.1 [**Mathematical Logic**]: Proof Theory; I.2.3 [**Artificial Intelligence**]: Deduction and Theorem Proving

## 1. INTRODUCTION

In this paper we propose a simple and natural multi-agent strategy logic, with explicit representations for individual and group strategies. The logic can be viewed as an extension of the well-known propositional logic of programs PDL. We show that the logic can express key notions of game theory and voting theory, such as Nash equilibrium, and the properties of voting rules that are used to prove the Gibbard-Satterthwaite theorem.

Unlike most other game logics, our logic uses explicit representations of group strategies in $N$-player games, with $N \geq 2$, and treats coalitions as a derived notion.

---

The logic we propose follows a suggestion made in Van Benthem [4] (in [11]) to apply the general perspective of action logic to reasoning about strategies in games, and links up to propositional dynamic logic (PDL), viewed as a general logic of action [29, 19]. Van Benthem takes individual strategies as basic actions and proposes to view group strategies as intersections of individual strategies (compare also [1] for this perspective). We will turn this around: we take the full group strategies (or: full strategy profiles) as basic, and construct individual strategies from these by means of strategy union.

A fragment of the logic we analyze in this paper was proposed in [10] as a logic for strategic reasoning in voting (the system in [10] does not have current strategies).

The plan of the paper is as follows. In Section 2 we review key concepts from strategic game theory, and hint at how these will show up in our logic. Section 3 does the same for voting theory. Section 4 gives a motivating example about coalition formation and strategic reasoning in voting. Section 5 presents the language of MASL, and gives the semantics. Next we show, in Section 6, that the key concepts of strategic game theory and voting theory are expressible in MASL. Section 7 extends the proof system for PDL to a sound and complete proof system for MASL. Section 8 gives an embedding of coalition logic into MASL. Section 9 extends MASL to an epistemic logic for reasoning about knowledge in games, Section 10 gives examples of what EMASL can express, and Section 11 sketches a calculus for EMASL. Section 12 concludes.

Key contributions of the paper are a demonstration of how PDL can be turned into a game logic for strategic games, and how this game logic can be extended to an epistemic game logic with PDL style modalities for game strategies and for epistemic operators. This makes all the logical and model checking tools for PDL available for analyzing properties of strategic games and epistemic strategic games.

## 2. GAME TERMINOLOGY

A strategic game form is a pair

$$(n, \{S_i\}_{i \in \{1,\ldots,n\}})$$

where $\{1, \ldots, n\}$ with $n > 1$ is the set of players, and each $S_i$ is a non-empty set of strategies (the available actions for player $i$). Below we will impose the restriction that the game forms are finite: each $S_i$ is a finite non-empty set.

We use $N$ for the set $\{1, \ldots, n\}$, and $S$ for $S_1 \times \cdots \times S_n$, and we call a member of $S$ a strategy profile. Thus, a strategy profile $s$ is an $n$-tuple of strategies, one for each player. If $s$ is a strategy profile, we use $s_i$ or $s[i]$ for its $i$-th component. Strategy profiles are in one-to-one correspondence to game outcomes, and in fact we can view $s \in S$ also as a game outcome [22].

Consider the prisoner's dilemma game PD for two players as an example. Both players have two strategies: $c$ for cooperate, $d$ for defect. The possible game outcomes are the four strategy profiles $(c,c), (c,d), (d,c), (d,d)$.

|   | c   | d   |
|---|-----|-----|
| c | c,c | c,d |
| d | d,c | d,d |

It is useful to be able to classify game outcomes. A $P$-outcome function for game form $(N, S)$ is a function $o : S \to P$.

For the example of the PD game, $o$ could be a function with range $\{x, y, z, u\}^2$, as follows:

|   | c   | d   |
|---|-----|-----|
| c | x,x | y,z |
| d | z,y | u,u |

If $C \subseteq N$, we let $S_C = \prod_{i \in C} S_i$ be the set of group strategies for $C$. If $s \in S_C$ and $t \in S_{N-C}$ we use $(s,t)$ for the strategy profile that results from combining $s$ and $t$, i.e., for the strategy profile $u$ given by

$$u[i] = s[i] \text{ if } i \in C, u[i] = t[i] \text{ otherwise.}$$

The group strategies for the PD game coincide with the strategy profiles.

An abstract game $G$ is a tuple

$$(N, S, \{\geq_i\}_{i \in N}),$$

where $(N, S)$ is a game structure, and each $\geq_i$ is a preference relation on $S_1 \times \cdots \times S_n$. These preference relations are assumed to be transitive, reflexive, and complete, where completeness means that for all different $s, t \in S$, one of $s \geq_i t$, $t \geq_i s$ holds.

In the PD game example, with the output function as above, the preferences could be fixed by adding the information that $z > x > u > y$.

The preference relations may also be encoded as numerical utilities. A payoff function or utility function for a player $i$ is a function $u_i$ from strategy profiles to real numbers. A payoff function $u_i$ represents the preference ordering $\geq_i$ of player $i$ if $s \geq_i t$ iff $u_i(s) \geq u_i(t)$, for all strategy profiles $s, t$.

A strategic game $G$ is a tuple

$$(N, \{S_i\}_{i \in N}, \{u_i\}_{i \in N})$$

where $N = \{1, \ldots, n\}$ and $u_i : S_1 \times \cdots \times S_n \to \mathbb{R}$ is the function that gives the payoff for player $i$. Aim of players in the game is to maximize their individual payoffs. We will use $u$ for the utility function, viewed as a payoff vector.

As an example, the PD game with payoffs as in the following picture, is a representation of the abstract version above.

|   | c   | d   |
|---|-----|-----|
| c | 2,2 | 0,3 |
| d | 3,0 | 1,1 |

It should be noted that payoff functions are a special case of output functions. In the example of PD with payoffs, we can view the payoff function as an output function with range $\{0, 1, 2, 3\}^2$.

Below, we will assume that output functions are of type $o : S \to P$, and we will introduce proposition letters to range over $P$. This allows us to view the game forms as modal frames, and the games including the output functions as models, with the output function fixing the valuation by means of "the valuation makes $p$ true in a state $s$ iff $s \in o^{-1}(p)$."

A special case of this is the case where the $P$ are payoff vectors. Valuations that are payoff vectors allow us to express preferences of the players for an outcome as boolean formulas (see below).

Let $(s'_i, s_{-i})$ be the strategy profile that is like $s$ for all players except $i$, but has $s_i$ replaced by $s'_i$. A strategy $s_i$ is a *best response* in $s$ if

$$\forall s'_i \in S_i \ u_i(s) \geq u_i(s'_i, s_{-i}).$$

A strategy profile $s$ is a (pure) Nash equilibrium if each $s_i$ is a best response in $s$:

$$\forall i \in N \ \forall s'_i \in S_i \ u_i(s) \geq u_i(s'_i, s_{-i}).$$

A game $G$ is *Nash* if $G$ has a (pure) Nash equilibrium.

These key notions of game theory will reappear below when we discuss the expressiveness of MASL.

## 3. VOTING AS A MULTI-AGENT GAME

Voting can be seen as a form of multi-agent decision making, with the voters as agents [14]. Voting is the process of selecting an item or a set of items from a finite set $A$ of alternatives, on the basis of the stated preferences of a set of voters. See [7] for a detailed account.

We assume that the preferences of a voter are represented by a ballot, where a ballot is a linear ordering of $A$. Let **ord**$(A)$ be the set of all ballots on $A$.

If there are three alternatives $a, b, c$, and a voter prefers $a$ over $b$ and $b$ over $c$, then her ballot is $abc$.

Assume the set of voters is $N = \{1, \ldots, n\}$. If we use $\mathbf{b}, \mathbf{b}'$ to range over ballots, then a profile $\mathbf{P}$ is a vector $(\mathbf{b}_1, \ldots, \mathbf{b}_n)$ of ballots, one for each voter. If $\mathbf{P}$ is a profile, we use $\mathbf{P}_i$ for the ballot of voter $i$ in $\mathbf{P}$.

The following represents the profile $\mathbf{P}$ where the first voter has ballot $abc$, the second voter has ballot $abc$, the third voter has ballot $bca$, and so on:

$$(abc, abc, bca, abc, cab, acb).$$

A *voting rule* $V$ for set of alternatives $A$ is a function from $A$-profiles to $\mathcal{P}^+(A)$ (the set of non-empty subsets of $A$). If $V(\mathbf{P}) = B$, then the members of $B$ are called the winners of $\mathbf{P}$ under $V$. A voting rule is *resolute* if $V(\mathbf{P})$ is a singleton for any profile $\mathbf{P}$.

Absolute majority is the voting rule that selects an alternative with more than 50 % of the votes as winner, and returns the whole set of alternatives otherwise. This is not the same as plurality, which selects an alternative that has the maximum number of votes as winner, regardless of whether more than half of the voters voted like this or not.

Strategizing is replacing a ballot $\mathbf{b}$ by a different one, $\mathbf{b}'$, in the hope or expectation to get a better outcome, where better is "closer to $\mathbf{b}$" in some sense. There are many ways to interpret 'better', and the particular choice does not matter. The way we will adopt (suggested in [32]) is to stipulate that $X$ is better than $Y$ if $X$ weakly dominates $Y$, that is, if every $x \in X$ is at least as good as every $y \in Y$ and some $x \in X$ is better than some $y \in Y$.

Formally: If $X, Y \subseteq A$ $X \neq \emptyset$, $Y \neq \emptyset$, and $\mathbf{b} \in \mathbf{ord}(A)$, then $X >_{\mathbf{b}} Y$ if $\forall x \in X \forall y \in Y$: $x = y$ or $x$ is above $y$ in $\mathbf{b}$, and $\exists x \in X \exists y \in Y$: $x$ is above $y$ in $\mathbf{b}$.

Let $\mathbf{P} \sim_i \mathbf{P}'$ express that profiles $\mathbf{P}$ and $\mathbf{P}'$ differ only in the ballot of voter $i$.

A voting rule is *strategy-proof* if $\mathbf{P} \sim_i \mathbf{P}'$ implies $V(\mathbf{P}) \geq_{\mathbf{b}} V(\mathbf{P}')$, where $\mathbf{b} = \mathbf{P}_i$ (so $\geq_{\mathbf{b}}$ expresses 'betterness' according to the $i$-ballot in $\mathbf{P}$).

To analyze voting as a game, think of casting an individual vote as a strategy. If we assume that the voting rule is fixed, this fixes the game outcome for each profile. The definition of 'betterness' determines the pay-off.

Player strategies are the votes the players can cast, so the set of individual strategies is the set $A$, for each player. Strategy profiles are the vectors of votes that are cast. Outcomes are determined by the voting rule; if the voting rule is resolute, outcomes are in $A$, otherwise in $\mathcal{P}^+(A)$. Preferences are determined by the voter types, plus some stipulation about how voters value sets of outcomes, given their type, in the case of non-resolute voting rules.

## 4. GROUP ACTION IN VOTING GAMES

To illustrate strategic reasoning and coalition formation in voting, we give an extended example. Suppose there are three voters $1, 2, 3$ and three alternatives $a, b, c$. Suppose the voting rule is plurality. Then each player or voter has the choice between actions $a$, $b$, and $c$.

Suppose 1 is the row player, 2 the column player, and 3 the table player. Then the voting outcomes are given by:

|    |   | a       | b       | c       |
|----|---|---------|---------|---------|
| a: | a | $a$     | $a$     | $a$     |
|    | b | $a$     | $b$     | $a,b,c$ |
|    | c | $a$     | $a,b,c$ | $c$     |

|    |   | a       | b | c       |
|----|---|---------|---|---------|
| b: | a | $a$     | $b$ | $a,b,c$ |
|    | b | $b$     | $b$ | $b$     |
|    | c | $a,b,c$ | $b$ | $c$     |

|    |   | a       | b       | c |
|----|---|---------|---------|---|
| c: | a | $a$     | $a,b,c$ | $c$ |
|    | b | $a,b,c$ | $b$     | $c$ |
|    | c | $c$     | $c$     | $c$ |

To determine the payoff function, we need information about the types of the voters. Suppose voter 1 has type (true ballot) $abc$. Then the betterness relation for 1 for the possible outcomes of the vote is given by:

$$a > b > c \text{ and } a > \{a, b, c\} > c.$$

Observe that neither $\{a,b,c\} > b$ nor $b > \{a,b,c\}$. So let's assume these give the same payoff, and fix the payoff function for voters of type $abc$ as

$$f(a) = 2, f(b) = f(\{a,b,c\}) = 1, f(c) = 0.$$

If we do similarly for the other voter types, then this fixes the strategic game for voting according to the plurality rule over the set of alternatives $\{a, b, c\}$.

So suppose 1 has ballot $abc$, 2 has ballot $bca$, and 3 has ballot $cab$. This gives the following strategic game form:

|    |   | a         | b         | c         |
|----|---|-----------|-----------|-----------|
| a: | a | (2, 0, 1) | (2, 0, 1) | (2, 0, 1) |
|    | b | (2, 0, 1) | (1, 2, 0) | (1, 1, 1) |
|    | c | (2, 0, 1) | (1, 1, 1) | (0, 1, 2) |

|    |   | a         | b         | c         |
|----|---|-----------|-----------|-----------|
| b: | a | (2, 0, 1) | (1, 2, 0) | (1, 1, 1) |
|    | b | (1, 2, 0) | (1, 2, 0) | (1, 2, 0) |
|    | c | (1, 1, 1) | (1, 2, 0) | (0, 1, 2) |

|    |   | a         | b         | c         |
|----|---|-----------|-----------|-----------|
| c: | a | (2, 0, 1) | (1, 1, 1) | (0, 1, 2) |
|    | b | (1, 1, 1) | (1, 2, 0) | (0, 1, 2) |
|    | c | (0, 1, 2) | (0, 1, 2) | (0, 1, 2) |

If the voters all cast their vote according to their true ballot, then 1 votes $a$, 2 votes $b$ and 3 votes $c$, and the outcome is a tie, $\{a, b, c\}$, with payoff $(1, 1, 1)$. This is a Nash equilibrium: the vote cast by each player is a best response in the strategy profile.

Now let's change the voting rule slightly, by switching to plurality voting with tie breaking, where $abc$ as the tie breaking order. This changes the plurality rule into a resolute voting rule. The new strategic game becomes:

|    |   | a         | b         | c         |
|----|---|-----------|-----------|-----------|
| a: | a | (2, 0, 1) | (2, 0, 1) | (2, 0, 1) |
|    | b | (2, 0, 1) | (1, 2, 0) | (2, 0, 1) |
|    | c | (2, 0, 1) | (2, 0, 1) | (0, 1, 2) |

|    |   | a         | b         | c         |
|----|---|-----------|-----------|-----------|
| b: | a | (2, 0, 1) | (1, 2, 0) | (2, 0, 1) |
|    | b | (1, 2, 0) | (1, 2, 0) | (1, 2, 0) |
|    | c | (2, 0, 1) | (1, 2, 0) | (0, 1, 2) |

|    |   | a         | b         | c         |
|----|---|-----------|-----------|-----------|
| c: | a | (2, 0, 1) | (2, 0, 1) | (0, 1, 2) |
|    | b | (2, 0, 1) | (1, 2, 0) | (0, 1, 2) |
|    | c | (0, 1, 2) | (0, 1, 2) | (0, 1, 2) |

If the players all vote according to their true preference, the outcome is $a$ because of the tie breaking, with payoff given by $(2, 0, 1)$. But this is no longer a Nash equilibrium, for player 2 can improve his payoff from 0 to 1 by casting vote $c$, which causes the outcome to change into $c$, with payoff $(0, 1, 2)$. The strategy triple $(a, c, c)$ is a Nash equilibrium.

So we are in a situation where the voting rule seems to favour voter 1 with ballot $abc$, because the tie breaking rule uses this order for tie breaking, and still the voter with this ballot ends up losing the game, because the other two players have an incentive to form a coalition against player 1.

## 5. A LANGUAGE FOR MASL

We will now turn to the description of strategic games like the PD game and the voting game in terms of actions in the spirit of PDL. We will take as our basic actions the full strategy profiles.

The reader is urged to think of a state in a game as a strategy vector where each player has determined her strategy. Strategy expressions in the MASL language are interpreted as relations on the space of all game states. Individual strategies emerge as unions of group strategies. An example is the strategy for the first player in the PD game to cooperate. This individual strategy is represented as $(c, ?\!?)$, and interpreted as in Figure 1.

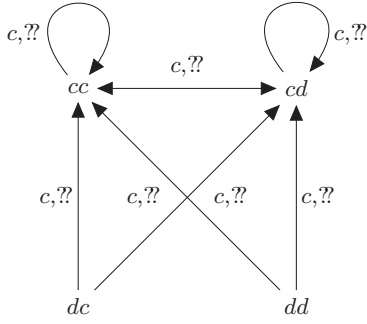

**Figure 1: Cooperation Strategy for Player 1 in PD Game**

Strategy terms of MASL are:

$$t_i \quad ::= \quad a \mid ?\!? \mid !\!!$$

Here $i$ ranges over the set of players $N$; and $a$ ranges over the set of all strategies $S_i$ for player $i$. A random term "$?\!?$" denotes an individual strategy for an adversary player, and "$!\!!$" denotes the current strategy of a player. Random terms serve to model what adversaries do, and current terms serve to model what happens when players stick to a previous choice.

As will become clear below, terms of the form $?\!?$ are used for succinctness; they could be dispensed with in favour of explicit enumerations of individual strategies.

From strategy terms we construct MASL strategy vectors, as follows:

$$\mathbf{c} \quad ::= \quad (t_1 \ldots, t_n)$$

The MASL strategy vectors occur as atoms and as modalities in MASL formulas. Allowing strategy terms as atomic formulas allows for succinct classification of game situations.

We assume that $p$ ranges over a set of game outcome values, that is: we assume an outcome function $o : S \to P$. The language is built in the usual PDL manner by mutual recursion of action expressions and formulas:

$$\phi \quad ::= \quad \top \mid \mathbf{c} \mid p \mid \neg \phi \mid \phi_1 \wedge \phi_2 \mid [\gamma]\phi$$
$$\gamma \quad ::= \quad \mathbf{c} \mid ?\phi \mid \gamma_1; \gamma_2 \mid \gamma_1 \cup \gamma_2 \mid \gamma^*$$

We will employ the usual abbreviations for $\bot$, $\phi_1 \vee \phi_2$, $\phi_1 \to \phi_2$, $\phi_1 \leftrightarrow \phi_2$ and $\langle \gamma \rangle \phi$.

Let $s \in S$. Then $s$ is a strategy profile, with individual strategies for player $i$ taken from $S_i$. We refer to the $i$-component of $s$ as $s[i]$. Thus, if $s = (a, b, b)$, then $s[1] = a$.

Let $i \in N$. Then $\llbracket \cdot \rrbracket^{S_i,s,i}$ is a function that maps each $t_i$ to a subset of $S_i$, and $\llbracket \cdot \rrbracket^{S,s}$ is a function that maps each strategy vector to a set of strategy profiles $\subseteq S$, as follows:

$$\llbracket a \rrbracket^{S_i,s,i} = \{a\}$$
$$\llbracket ?\!? \rrbracket^{S_i,s,i} = S_i$$
$$\llbracket !\!! \rrbracket^{S_i,s,i} = \{s[i]\}$$
$$\llbracket (t_1 \ldots, t_n) \rrbracket^{S,s} = \llbracket t_1 \rrbracket^{S_1,s,1} \times \cdots \times \llbracket t_n \rrbracket^{S_n,s,n}$$

For example, let the set of individual strategies for each player be $A = \{a, b, c\}$, and let $n = 3$ (as in the voting example in Section 4). Then a strategic change by the first player to $b$, while both other players stick to their vote is expressed as $(b, !\!!, !\!!)$. In a game state $(a, b, b)$ this is interpreted as $\{((a,b,b),(b,b,b))\}$.

A strategic change by the first player to $b$, given that the second player sticks to her vote, while the third player may or may not change, is expressed by $(b, !\!!, ?\!?)$. In the context of a strategy profile $s = (a, b, c)$, this is interpreted as follows:

$$\llbracket (b, !\!!, ?\!?) \rrbracket^{A,s} = \{b\} \times \{b\} \times \{a, b, c\}.$$

$(?\!?, c, c)$ represents the group strategy where players 2 and 3 both play $c$. This is a strategy for the coalition of 2 and 3 against 1.

The formula that expresses that the coalition of 2 and 3 can force outcome $c$ by both voting $c$ is (abbreviating the singleton outcome $\{c\}$ as $c$):

$$[(?\!?, c, c)]c.$$

The strategy $(?\!?, ?\!?, c)$ is different from $(!\!!, !\!!, c)$, for the latter expresses the individual strategy for player 3 of playing $c$, in a context where the two other players do not change their strategy.

The relational interpretation for coalition strategies follows the recipe proposed in [4], but with a twist. We interpret a strategy for an individual player as a relation on a set of game states, by taking the union of all full strategy relations that agree with the individual strategy. So the strategies for the individual players are choices that emerge from taking unions of vectors that determine the game outcome completely. If we assume that the players move together, without information about moves of the other players, then the individual strategies are choices, but an individual choice does not determine an outcome. Only the joint set of all choices does determine an outcome.

So if we represent a strategy for player $i$ as a relation, then we have to take into account that the individual choice of $i$ does need information about how the others move to determine the outcome. The relation for the individual choice $a$ of player $i$ is given by

$$\llbracket (?\!?, \cdots, ?\!?, a, ?\!?, \cdots, ?\!?) \rrbracket^{S,s}$$
$$= S_1 \times \cdots \times S_{i-1} \times \{a\} \times S_{i+1} \times \cdots \times S_n.$$

This relation is computed from all choices that the other players could make (all strategies for the other players).

Compare this with

$$\llbracket (!\!!, \cdots, !\!!, a, !\!!, \cdots, !\!!) \rrbracket^{S,s} =$$
$$\{s[1]\} \times \cdots \{s[i-1]\} \times \{a\} \times \{s[i+1]\} \times \cdots \{s[n]\}.$$

This is the action where player $i$ switches to $a$, while all other players stick to their strategies.

The picture in Figure 2 gives the interpretation of the $(c, !\!!)$ strategy vector in the PD game.

This generalizes to coalitions, as follows. A strategy for a coalition is a choice for each of the coalition members, and the corresponding relation is the union of all full strategy relations that agree with the coalition strategy. Compare the definition of the $\llbracket \cdot \rrbracket^{S,s}$ function for strategy vectors above.

This gives an obvious recipe for turning strategic game forms with outcome functions into Kripke models. Let $(N, S)$

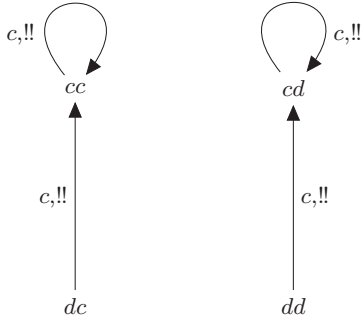

**Figure 2: Interpretation of $(c, !!)$ in PD Game**

be a strategic game form and let $o : S \to P$ be an outcome function, and let $s \in S$ be a strategy profile.

Then the truth definition for MASL, with respect to $M = (N, S, o)$ and $s$ is given by:

$$
\begin{aligned}
M, s &\models \top && \text{always} \\
M, s &\models \mathbf{c} && \text{iff} \quad s \in [\![\mathbf{c}]\!]^{S,s} \\
M, s &\models p && \text{iff} \quad s \in o^{-1}(p) \\
M, s &\models \neg \phi && \text{iff} \quad M, s \not\models \phi \\
M, s &\models \phi_1 \wedge \phi_2 && \text{iff} \quad M, s \models \phi_1 \text{ and } M, s \models \phi_2 \\
M, s &\models [\gamma]\phi && \text{iff} \quad \text{for all } t \text{ with } (s,t) \in [\![\gamma]\!]^M : \\
& && \quad M, t \models \phi
\end{aligned}
$$

$$
\begin{aligned}
{[\![\mathbf{c}]\!]}^M &= \{(s,t) \mid t \in [\![\mathbf{c}]\!]^{S,s}\} \\
{[\![?\phi]\!]}^M &= \{(s,s) \mid M, s \models \phi\} \\
{[\![\gamma_1;\gamma_2]\!]}^M &= [\![\gamma_1]\!]^M \circ [\![\gamma_2]\!]^M \\
{[\![\gamma_1 \cup \gamma_2]\!]}^M &= [\![\gamma_1]\!]^M \cup [\![\gamma_2]\!]^M \\
{[\![\gamma^*]\!]}^M &= ([\![\gamma]\!]^M)^*,
\end{aligned}
$$

where $\circ$ is used for relation composition, and $*$ for reflexive transitive closure.

Note that it is assumed that the signature of the language matches that of the model: for interpretation in $M = (N, S, o)$ with $o : S \to P$, we assume that strategy vectors of the language have length $n$, that the terms $t_i$ of the language get interpreted as subsets of $S_i$, and that the propositionional atoms range over $P$.

## 6. EXPRESSIVENESS OF MASL

We give examples to demonstrate that MASL expresses key concepts of game theory, voting theory, social choice theory and iterated game playing, in a natural way.

*Abbreviations.*
Let $(i_a, \overline{!!})$ abbreviate the strategy vector

$$(!!, \cdots, !!, a, !!, \cdots, !!),$$

with $a$ in $i$-th position, and $!!$ everywhere else.

Using this, let $[(i, \overline{!!})]\phi$ abbreviate $\bigwedge_{a \in S_i}[(i_a, \overline{!!})]\phi$. Then $[(i, \overline{!!})]\phi$ expresses that all strategies to which player $i$ can switch from the current strategy profile result in a strategy profile where $\phi$ holds (provided that the other players keep their strategies fixed).

Let $(i_a, \overline{??})$ abbreviate the strategy vector

$$(??, \cdots, ??, a, ??, \cdots, ??),$$

with $a$ in $i$-th position, and $??$ everywhere else.

Using this, let $[(i, \overline{??})]\phi$ abbreviate $\bigwedge_{a \in S_i}[(i_a, \overline{??})]\phi$. Then $[(i, \overline{??})]\phi$ expresses that all strategies for $i$ guarantee $\phi$, no matter what the other players do.

Let $(\overline{??})$ abbreviate $(??, \cdots, ??)$ (the strategy vector that everywhere has $??$). Then $\langle(\overline{??})\rangle\phi$ expresses that in some game state $\phi$ holds.

*Representing Payoffs.*
To represent payoffs, we will assume that basic propositions are payoff vectors $u$, and that the payoff values are in a finite set $U$ (the set of all utilities that can be assigned in the game). Next, define $u_i \geq v$ as $\bigvee_{w \in U, w \geq v} u[i] = w$ and $u_i > v$ as $\bigvee_{w \in U, w > v} u[i] = w$. Then $u_i \geq v$ expresses that player $i$ gets at least $v$, and $u_i > v$ expresses that player $i$ gets more than $v$ (compare [34] for a similar approach).

*Weak Dominance.*
Using the above abbreviations, we can express what it means for an $i$-strategy $a$ to be *weakly dominant*. Intuitively, it means that $a$ is as least as good for $i$ against any moves the other players can make as any alternative $b$ for $a$. In our logic:

$$\bigwedge_{v \in U} \bigwedge_{b \in A - \{a\}} [(i_b, \overline{??})](u_i \geq v \to \langle(i_a, \overline{!!})\rangle u_i \geq v).$$

*Nash Equilibrium.*
The following formula expresses that the current strategy profile is a Nash equilibrium:

$$\bigwedge_{i \in N} \bigvee_{v \in U} (u_i \geq v \wedge [(i, \overline{!!})]\neg u_i > v).$$

The following formula expresses that the game is Nash:

$$\langle(\overline{??})\rangle \bigwedge_{i \in N} \bigvee_{v \in U} (u_i \geq v \wedge [(i, \overline{!!})]\neg u_i > v).$$

*Plurality Voting.*
For the application of MASL to voting, assume the output function produces an ordered pair consisting of the outcome of the voting rule for a profile, plus the utility vector for the players for that profile.

Let $A$ be the set of alternatives. Let $P_a$ be the set of all full strategy vectors where $a$ gets more votes than any other alternative. Then

$$\bigwedge_{x \in A} \bigwedge_{\mathbf{c} \in P_x} [\mathbf{c}]x$$

expresses that the game is a voting game with plurality rule. This is easily extended to a formula that expresses the rule of plurality voting with tie breaking.

*Resoluteness.*
Assume that the proposition $a$ expresses that $a$ is among the winners given the current profile. A voting rule is *resolute* if there is always exactly one winner. Viewing voting according to a voting rule as a game, the following formula expresses that the game is resolute:

$$[(\overline{?})] \bigvee_{a \in A} (a \wedge \bigwedge_{b \in A-\{a\}} \neg b).$$

*Strategy-Proofness.*

A voting rule is *strategy proof* if it holds for any profile $S$ and for any player (voter) $i$ that changing his vote (action) does not give an outcome that is better (according to the preferences of $i$ in $S$) that the outcome in $S$. This is expressed by the following formula:

$$[(\overline{?})] \bigwedge_{i \in N} \bigvee_{v \in U} (u_i \geq v \wedge \neg \langle (i, \overline{!}) \rangle u_i > v).$$

*Non-Imposedness.*

A voting rule is (weakly) *non-imposed* if at least three outcomes are possible. Viewing voting as a game, we can use the following formula to express this:

$$\bigvee_{a \in A} \bigvee_{b \in A-\{a\}} \bigvee_{c \in A-\{a,b\}} (\langle (\overline{?}) \rangle a \wedge \langle (\overline{?}) \rangle b \wedge \langle (\overline{?}) \rangle c).$$

*Dictatorship.*

In a multi-agent game setting, a dictator is a player who can always get what he wants, where getting what you want is getting a payoff that is at least as good as anything any other player can achieve. Here is the formula for that, using the abbreviation $(i, \overline{?})$:

$$\bigvee_{v \in U} \bigwedge_{j \in N-\{i\}} [(?)](\neg u_j > v \wedge \langle (i, !!) \rangle u_i \geq v).$$

*Gibbard-Satterthwaite.*

The classic Gibbard-Satterthwaite theorem [15, 30] states that all reasonable voting rules allow strategizing, or put otherwise, that no reasonable voting rule is strategy-proof.

Resoluteness, strategy-proofness, non-imposedness and dictatorship are the four properties in terms of which the Gibbard-Satterthwaite theorem is formulated, and in fact, the theorem can be stated and proved in our logic. What the theorem says semantically is:

$$\text{Res}, \text{SP}, \text{NI} \models \text{Dict}.$$

It follows from the completeness of the logic (Section 7 below) that for every choice of MASL language (where the choice of language fixes the number of players/voters $N$ and the set of alternatives $A$, with $|A| > 3$), the following can be proved:

$$\text{Res}, \text{SP}, \text{NI} \vdash \text{Dict}.$$

*Meta-Strategies: Tit-for-Tat.*

Tit-for-tat as a meta-strategy for the PD game [3] is the instruction to copy one's opponents last choice, thereby giving immediate, and rancour-free, reward and punishment. Figure 3 gives a picture of the tit-for-tat meta-strategy for player 2, with the states indicating the outcomes of the last play of the game.

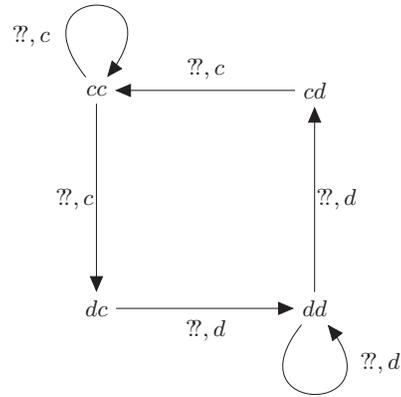

**Figure 3: Tit-for-tat Meta-Strategy for Player 2 in PD Game**

This works because we may think of the current state of the game as the result of the last play of PD, remembered in the state. Testing the state yields the clue for whether the reward action $(?, c)$ or the punishment action $(?, d)$ has to be executed. Thus, the following MASL action expression describes this meta-strategy for player 2.

$$(?(c, ?); (?, c) \cup ?(d, ?); (?, d))^*$$

What this says is: if the last action by the opponent was a $c$, then reward, otherwise (the last action by the opponent was a $d$) punish. To turn this into a meta-strategy for player 1, just swap all pairs:

$$(?(?, c); (c, ?) \cup ?(?, d); (d, ?))^*$$

Note that tit-for-tat for the PD game boils down to the same thing as the *copycat* meta-strategy, where a player always copies the last move of the opponent. So the player 1 copycat and player 2 copycat meta-strategies for the PD game are also given by the above strategy expressions.

## 7. A COMPLETE CALCULUS FOR MASL

To axiomatize this logic, we can use the well-known proof system for PDL [31, 23], with appropriate axioms for the strategy vectors added.

Call a strategy vector $\mathbf{c} = (t_1 \ldots, t_n)$ *determined* if for no $i \in N$ it is the case that $t_i = ?$.

Vector axioms are:

1. Effectivity:

$$[\mathbf{c}]\mathbf{c}.$$

2. Seriality:

$$\langle \mathbf{c} \rangle \top.$$

3. Functionality:

$$\langle \mathbf{c} \rangle \phi \rightarrow [\mathbf{c}]\phi$$

for all determined strategy vectors $\mathbf{c}$.

4. Adversary power:

Let **c** have ?? in position $i$, and let $\mathbf{c}_a^i$ be the result of replacing ?? in position $i$ in **c** by $a$. Then:

$$[\mathbf{c}]\phi \leftrightarrow \bigwedge_{a \in S_i} [\mathbf{c}_a^i]\phi.$$

Note that this uses the assumption that the set $S_i$ of available actions for player $i$ s finite.

5. Determinate current choice:

   Let **c** have !! in position $i$, and let $\mathbf{c}_a^i$ be the result of replacing !! at position $i$ in **c** by $a$. Then:

   $$(i_a, \overline{!!}) \rightarrow (\mathbf{c} \leftrightarrow \mathbf{c}_a^i).$$

The effectivity axiom says that execution of a strategy vector always makes the vector true.

The seriality axiom says that every strategy vector can be executed.

The functionality axiom says that determined strategy vectors are functional. This expresses that the outcome is determined if every player makes a determinate choice. This axiom does not hold for vectors that are not determined. The vector $(c, ??)$ has $|S_2|$ possible outcomes.

The adversary power axiom spells out what an adversary player can do. This defines the meaning of ?? terms.

The determinate current choice axiom fixes the meaning of !! terms.

These axioms are sound for the intended interpretation. Completeness can be shown by the usual canonical model construction for PDL (see [19, 6]):

THEOREM 1. *The calculus for MASL is complete.*

MASL has the same complexity for model checking and satisfiability as PDL: Model checking for PDL and MASL is PTIME-complete [20]. Sat solving for PDL and MASL is EXPTIME-complete [6]. Model checking for formulas that use only the modal fragment of MASL (modalities without Kleene star) can be done more efficiently, e.g., by using the algorithm of [13] that runs in time $O(|M| \times |\phi|)$.

The important thing to note is that the standard model checking tools for modal logic and PDL can now be used for strategic games, using the MASL extension of PDL.

## 8. CONNECTION TO COALITION LOGIC

Our approach links directly to coalition logic [27] (see also [4] for this connection). Coalition logic has the following syntax:

$$\phi ::= \top \mid p \mid \neg\phi \mid \phi \wedge \phi \mid [C]\phi$$

where $p$ ranges over basic propositions, and $C \subseteq N$, with $N$ the set of agents. Intended meaning of $[C]\phi$ is that the coalition $C$ is able to force the game outcome to be in $\phi$.

Again, let $(N, S)$ be a strategic game form, let $o : S \rightarrow P$ be an outcome function, and let $s \in S$ be a strategy profile. Assume models $M$ of the form $(N, S, o)$. Coalition logic, the way it is presented in [27], is a bit mysterious about how valuations enter into game forms, but we can fix this by using the output functions in the same manner as in the semantics of MASL. Formulas of coalition logic are interpreted in strategy profiles $s$ of $M$, as follows.

Recall that $S_C$ is the set of group strategy functions for $C$, and that if $s \in S_C$ and $t \in S_{N-C}$, then $(s, t)$ is the strategy profile where members of $C$ choose according to $s$ and all others choose according to $t$.

$$\begin{aligned} M, s &\models p & \text{iff} \quad & s \in o^{-1}(p). \\ M, s &\models \neg\phi & \text{iff} \quad & M, s \not\models \phi. \\ M, s &\models \phi_1 \wedge \phi_2 & \text{iff} \quad & M, s \models \phi_1 \text{ and } M, s \models \phi_2. \\ M, s &\models [C]\phi & \text{iff} \quad & \exists t \in S_C \forall u \in S_{N-C} \\ & & & M, (t, u) \models \phi. \end{aligned}$$

Let $\dot{C}$ be the set of all strategies for coalition $C$ against all other players.

If we assume that for each player $i$ the set $S_i$ of possible strategies for $i$ is finite, then $\dot{C}$ is finite as well, and $\dot{C}$ is defined by

$$\{(t_1, \ldots, t_n) \mid t_i \in S_i \text{ if } i \in C, t_i = ?? \text{ otherwise }\}.$$

This means we can construct the formula

$$\bigvee_{\mathbf{c} \in \dot{C}} [\mathbf{c}]\phi.$$

The translation instruction $Tr$ for turning coalition logic into MASL becomes:

$$\begin{aligned} \mathrm{Tr}(p) &:= p \\ \mathrm{Tr}(\neg\phi) &:= \neg\mathrm{Tr}(\phi) \\ \mathrm{Tr}(\phi_1 \wedge \phi_2) &:= \mathrm{Tr}(\phi_1) \wedge \mathrm{Tr}(\phi_2) \\ \mathrm{Tr}([C]\phi) &:= \bigvee_{\mathbf{c} \in \dot{C}} [\mathbf{c}]\mathrm{Tr}(\phi). \end{aligned}$$

Induction on formula structure now proves:

THEOREM 2. $M, s \models_{CL} \phi$ iff $M, s \models_{MASL} Tr(\phi)$.

This assumes that the set of strategies for each agent is finite, as this is a basic assumption of MASL. This finiteness restriction aside, the main difference between coalition logic and MASL is that MASL is explicit about coalition strategies where coalition logic is not. Many key concepts of strategic game theory and voting theory that MASL can express are beyond the reach of coalition logic.

## 9. EPISTEMIC MASL

MASL uses PDL as an action logic for game actions. It is well-known that PDL also can be given an epistemic interpretation [5]. The language of Epistemic Multi Agent Strategy Logic (EMASL) combines the strategy interpretation of PDL with the epistemic interpretation of PDL. For that, a new set of PDL actions is thrown in, but this time with an epistemic/doxastic interpretation. Here is the extended language:

$$\begin{aligned} \phi &::= \top \mid \mathbf{c} \mid p \mid \neg\phi \mid \phi_1 \wedge \phi_2 \mid [\gamma]\phi \mid [\alpha]\phi \\ \gamma &::= \mathbf{c} \mid ?\phi \mid \gamma_1; \gamma_2 \mid \gamma_1 \cup \gamma_2 \mid \gamma^* \\ \alpha &::= i \mid i\breve{} \mid ?\phi \mid \alpha_1; \alpha_2 \mid \alpha_1 \cup \alpha_2 \mid \alpha^* \end{aligned}$$

$i$ ranges over the set $N$ of agents. $i\breve{}$ denotes the converse of the $i$ relation.

The interpretations of the $i$ operators (the atoms of $\alpha$ actions) can be arbitrary. Define **i** as $(i \cup i\breve{})^*$, and you have a reflexive, symmetric and transitive knowledge operator (see [12]).

Note that tests appear both in the action expressions and in the epistemic expressions. Thus, actions can be conditioned by knowledge, and knowledge can refer to action. This allows the representation of strategies like "If I know that playing $a$ results in $\phi$, then play $a$, else play $b$" (action conditioned by knowledge), and the representation of epistemic relations expressing what will become known as a result of a certain strategy (knowledge referring to action).

To interpret this language, we define *intensional game forms* from (extensional) game forms. An intensional game form is a tuple

$$(N, W, R_1, \ldots, R_n)$$

where

- $W$ is a set of pairs $(G, s)$ where $G = (N, S)$ is a game form with $s \in S$,
- each $R_i$ is a binary relation on $W$.

These intensional game forms can be viewed as Kripke frames. As before, they can be turned into models by using an output function $o : S \to P$ to define the valuation. For that, extend $o$ to $W$ by means of the stipulation saying that the output of a game-profile pair is determined by its profile component:

$$o(G', s') = o(s').$$

Since $s' \in S' \subseteq S$ for each $S'$, this is well-defined.

Let $M$ be an intensional game form $(N, W, R_1, \ldots, R_n)$ based on $G = (N, S)$ and let $o : W \to P$ be an output function that is extended from an output function $o : S \to P$ for $G$. Let $w \in W$.

Then the truth definition of EMASL formulas in $M, w$ is given by (only clauses that differ from the MASL version shown):

$$M, w \models \mathbf{c} \quad \text{iff} \quad w = ((N, S), s) \text{ and } s \in [\![\mathbf{c}]\!]^{S,s}$$
$$M, w \models [\alpha]\phi \quad \text{iff} \quad \text{for all } w' \text{ with } (w, w') \in [\![\alpha]\!]^M: \\ M, w' \models \phi$$

$$[\![\mathbf{c}]\!]^M = \{(w, w') \mid w = ((N, S), s), w' = ((N, S), t) \\ \text{with } t \in [\![\mathbf{c}]\!]^{S,s}\}$$
$$[\![\mathbf{i}]\!]^M = R_i$$
$$[\![\mathbf{\breve{i}}]\!]^M = (R_i)^{\breve{}}$$

One way to base an intensional game form on a game form $G = (N, S)$ is by putting $W = \{(G, s) \mid s \in S\}$ and

$$R_i = \{((G, s), (G, s')) \mid s[i] = s'[i]\}$$

for all $i \in N$. Call this the *epistemic lift* of $G$, and denote it with $G^\#$.

Then in $G^\#$ the accessibility relations express that every player can distinguish between her own actions, but not between those of other players.

For the PD game, this gives a model where every player knows her move, and the two possible strategies for her opponent. Furthermore, it is common knowledge that there is no coordination between the actions of the two players: the relation $(R_1 \cup R_2)^*$, denoted by the EMASL expression $(\mathbf{1} \cup \mathbf{2})^*$, is the whole set of strategy profiles.

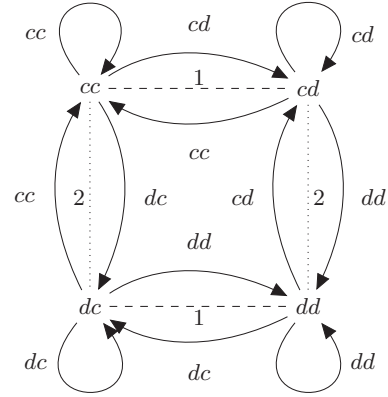

**Figure 4: Epistemic PD Game Form**

This is pictured in Figure 4, with dashed lines for the accessibilities of player 1, dotted lines for those of player 2, and reflexive epistemic arrows omitted.

In epistemic lifts of game forms it is common knowledge among all players what is the nature of the game; more in particular it is common knowledge what are the available strategic options for all players.

This assumption that the nature of the game is common knowledge is dropped for intensional game forms that are built by means of *strategy restrictions* from an (extensional) game form.

Let $G' \sqsubseteq G$ if $G = (N, \{S_i \mid i \in N\})$, $G' = (N, \{S'_i \mid i \in N\})$, and for all $i \in N$: $S'_i \subseteq S_i$. Call $G'$ a strategy restriction of $G$.

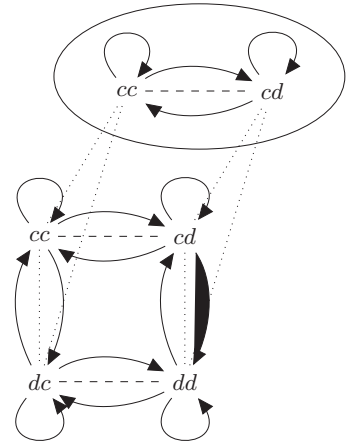

**Figure 5: Restriction in PD Game.**

An intensional game form built from strategy restriction from PD is given in Figure 5. This pictures a situation where the first player is committed to $c$, but the other player does not know this. The oval indicates the actual game; this is confused by player 2 with the full PD game (dotted lines for the accessibility relation of player 2).

## 10. EXPRESSIVENESS OF EMASL

EMASL extends MASL, so every concept from game theory, voting theory and social choice theory that is expressible in MASL is expressible in EMASL. Many concepts from social choice theory have epistemic versions. Here is one example.

*Knowing Dictatorship.*

A dictator in a multi agent game was defined as a player who is always able to get the best deal. A knowing dictator is a player who not only has this ability, but also *knows* that he has it:

$$[\mathbf{i}] \bigvee_{v \in U} \bigwedge_{j \in N-\{i\}} [(??)](\neg u_j > v \land \langle(i, !!)\rangle u_i \geq v).$$

As an example, consider player 2 in the game pictured in Figure 5, with an output function giving appropriate utilities for the PD game. Player 2 is a dictator for this game, for he can force the outcome $cd$, with best payoff for 2. But player 2 is not aware of this fact: for all he knows, he could end up in state $dd$, with worse payoff for him than $cd$.

*Gibbard-Satterthwaite, Epistemically.*

Resoluteness, strategy-proofness, non-imposedness, dictatorship and knowing dictatorship are all expressible in EMASL. Here is a new type of question. Consider the class of epistemic lifts $G^\#$ of strategic game forms $G$ based on resolute, strategy-proof and non-imposed voting rules. Then the MASL proof of the GS theorem lifts to EMASL, so every such game has a dictator. But does every such game also have a knowing dictator? What are the minimum epistemic conditions to make the epistemic GS theorem go through in intensional games? It also make sense to formulate an epistemic version of strategy-proofness, stating that players do not know that they can improve their payoff by voting strategically. This is a weakening of strategy-proofness, and we can investigate under which epistemic conditions it is enough to derive GS, or to derive epistemic GS.

## 11. A CALCULUS FOR EMASL

There are various classes of intensional game models that one might want to axiomatize. As an example, we consider the class of epistemic lift models $(G^\#, o)$, where $G = (N, S)$ is a finite strategic game form and $o : S \to P$ is an output function for $G$.

Notice that the axioms of MASL are sound for this class, so that we can extend the calculus for MASL, to get a calculus for reasoning about epistemic lift models, as follows.

- Propositional axioms, modus ponens, necessitation for $\gamma$ and $\alpha$.
- PDL axioms for $\gamma$ modalities.
- PDL axioms for $\alpha$ modalities.
- The five MASL vector axioms.
- $\phi \to [i]\langle \breve{i} \rangle \phi$.
- $\phi \to [\breve{i}]\langle i \rangle \phi$.
- $[(i_a, !!)][\mathbf{i}](i_a, !!)$.
- $\bigwedge_{j \in N-\{i\}} [(j_a, !!)] \neg [\mathbf{i}](j_a, !!)$.

The two axioms for $\breve{i}$ are the standard modal axioms for converse. The first axiom for $[\mathbf{i}]$ expresses that player $i$ can distinguish between his own actions, and the second axiom for $[\mathbf{i}]$ expresses that $i$ cannot distinguish between the actions of other players. This gives a sound and complete system for reasoning about epistemic lift models.

## 12. RELATED AND FURTHER WORK

The present approach is closest to [4], to which it is indebted. Instead of constructing group strategies from individual strategies by relation intersection, we take complete group strategies as basic in the semantics, and construct strategies for subgroups and individuals by relation union.

Strategic reasoning is related to multimodal logic $K_n$ with intersection in [1], where group strategies are constructed from individual strategies by means of relation intersection. Strategies are not explicit, and many key notions from game theory are not expressible.

The present approach is close to [18, 21] where PDL is taken as a starting point to formulate expressive STIT logics for analzying agency in games. In [33] a special purpose logic for reasoning about social choice functions is proposed, with an analysis to the concept of strategy-proofness, in terms of a modality for expressing that certain players stick to their current choice.

Coalition logic is a close kin of Alternating-time Temporal Logic [2, 16]. This has various extensions, of which CATL [34] deals explicitly with strategic reasoning. An important difference with the present approach is that ATL and CATL focus on extensive rather than strategic games.

Strategic reasoning is also the topic of game logics such as [24, 25, 28]. These logics focus on the theory of two-player games, and also use the regular operations for strategy construction. The imporant difference is that in game logic the regular operations are applied to single player strategies. We hope to study the connection with game logic and with game algebra [17, 35] more precisely in future work.

Our work provides a framework for extending the exploration of knowledge-theoretic properties of strategic voting in [8]. In [26] the notion of knowledge manipulation in games is discussed, which is in the compass of EMASL, as is the analysis of the role of knowledge and ignorance in voting manipulation in [9].

There are two important limitations of MASL and EMASL, in the versions presented here: the restriction to finite ranges of individual actions, and the restriction to strategic games.

The first restriction could be lifted by extending the language with quantification over $i$-strategies. To lift the second restriction, one could introduce a register for keeping track of the players that have made their move. This would get us closer to ATL, for such a register can be viewed as a clock. It remains to be seen whether this extension could still be handled naturally by a PDL-based approach.

In [5], epistemic PDL is used as a base system to which operators for communication and change are added. In future work, we hope to extend the framework of EMASL in a similar way with communication and changes operators. Communications change the epistemics of the game by informing players about strategies of other players. Change operations change the game by changing outcomes or utilities. The switch from plurality voting to plurality voting with tie breaking could be modelled as such a change.

*Acknowledgements*

This paper has benefitted from conversations with Johan van Benthem, Ulle Endriss, Floor Sietsma and Sunil Simon. We also wish to thank four anonymous TARK reviewers for comments and suggestions.